\documentstyle[12pt,fleqn]{article}
\begin{document}
\sloppy
\parindent 8mm
\mathindent 8mm

\hspace*{\fill}TECHNION-PHYS-94-LP2
\vspace*{2cm}
\begin{center}
{\large Quantum Field Theory with Null-Fronted Metrics}

\bigskip Orit Levin and Asher Peres

\bigskip{\sl Department of Physics, Technion---Israel Institute of
Technology, 32 000 Haifa, Israel}

\end{center}\vspace{2cm}

There is a large class of classical null-fronted metrics in which a free
scalar field has an infinite number of conservation laws. In particular,
if the scalar field is quantized, the number of particles is conserved.
However, with more general null-fronted metrics, field quantization
cannot be interpreted in terms of particle creation and annihilation
operators, and the physical meaning of the theory becomes
obscure.\vspace{2cm}

\noindent 1994 PACS numbers: \ 04.62.+v \ 04.20.Cv \ 03.50.-z\vfill

\noindent\hrulefill\\
Please address all correspondence to the second author.\\
E-mail: peres@techunix.technion.ac.il\\
Fax: + 972 4 221514\newpage

Conventional canonical dynamics, and likewise quantum field theory, use
a time-like coordinate as the evolution parameter. However, it was
pointed out long ago by Dirac [1] that it may be advantageous to use a
null coordinate for that purpose. The mathematical consistency of
quantum field theory on null slabs in Minkowski space was formally
proved by Leutwyler, Klauder, and Streit [2]. This approach has led to
numerous applications in particle physics~[3].

In this work, we examine the consistency of canonical (or quantum) field
theory when a null coordinate is used as the evolution parameter in
curved spacetime. This choice may be advantageous for additional
reasons: in some cases, a complete foliation of spacetime may be
possible with null hypersurfaces, while it is impossible with space-like
hypersurfaces. We start by giving a very simple derivation of a result
which has been known for some time: classical plane-fronted
gravitational waves do not polarize the vacuum; namely, their
interaction with a free quantized field does not create particles
[4--6]. We show that this property is a special case of an infinity of
conservation laws, valid for any metric which obeys the conditions
\begin{equation}
g^{00}=g^{01}=g^{02}=0,
\end{equation}
and
\begin{equation}
\partial g^{\alpha\beta}/\partial x^3=0.
\end{equation}
However, when we consider more general metrics, we run into severe
consistency problems. In particular, a free quantum field cannot be
interpreted as consisting of particle creation and annihilation
operators.

The condition $g^{00}=0$ means that hypersurfaces with constant
$x^0\equiv u$ (which will be used as ``time'' in the following
discussion), have a null normal vector $\partial_\alpha u$. This is why
$u$ is called a ``null coordinate.'' Equation~(2) is the definition of
the ``null-fronted'' symmetry. We shall use numerical indices 0, 1, 2,
3, to denote the coordinates $u, x, y, v$, respectively. Greek indices
refer to all four coordinates, and Latin indices to coordinates 1 and 2
only. Plane-fronted gravitational waves~[7] are a special case of
null-fronted metric: the only nonvanishing components of
$g^{\alpha\beta}$ are \,$g^{03}=-g^{11}=-g^{22}=1$, and $g^{33}$ which
satisfies $\,\sum_m\partial^2g^{33}/(\partial x^m)^2=0$. Null-fronted
metrics with more general values of $g^{\alpha\beta}$ indicate the
presence of a nonvanishing energy-momentum tensor, which may sometimes
(but not always) be physically realizable [8].

Equation (2), together with the geodesic equation,
\begin{equation}
\frac{d}{ds}\left(g_{\lambda\mu}\,\frac{dx^\mu}{ds}\right)=
\frac{1}{2}\;\frac{\partial g_{\alpha\beta}}{\partial x^\lambda}\;
\frac{dx^\alpha}{ds}\;\frac{dx^\beta}{ds}\,,
\end{equation}
which describes the motion of a free classical particle, implies that
the momentum component,
\begin{equation}
p_3\equiv m\,g_{3\alpha}\,(dx^\alpha/ds),
\end{equation}
is a constant of the motion ($m$ is the mass of the particle). Since
{\em all\/} classical orbits have this property, we expect the existence
of infinitely many conservation laws in a free field theory (classical
or quantal) having a null-fronted background metric.

Consider indeed a free classical real scalar field $\phi$, satisfying
the Klein-Gordon equation
$\,(\gamma^{\alpha\beta}\,\phi_{,\alpha})_{,\beta}=-\kappa^2\gamma\phi$,
where \,$\gamma\equiv(-{\rm Det}\,g_{\alpha\beta})^{1/2}$ \,and
$\,\gamma^{\alpha\beta}\equiv\gamma\,g^{\alpha\beta}$. (The
$\gamma^{\alpha\beta}$ are denoted by Gothic letters in the older
literature.) All these expressions are functions of $u$, $x$, and $y$
only. Explicitly, the Klein-Gordon equation is
\begin{equation}
2\,\gamma^{03}\,\phi_{,03}+\gamma^{03}{}_{,0}\,\phi_{,3}+
\gamma^{33}\,\phi_{,33}+
2\,\gamma^{m3}\,\phi_{,m3}+\gamma^{m3}{}_{,m}\,\phi_{,3}+
(\gamma^{mn}\,\phi_{,m})_{,n}=-\kappa^2\,\gamma\,\phi.\label{KG}
\end{equation}
Let us perform a Fourier transformation with respect to $v$ (only):
\begin{equation}
\chi(u,x,y,p)\equiv
(2\pi)^{-1/2}\int^\infty_{-\infty}e^{ivp}\,\phi(u,x,y,v)\,dv.
\end{equation}
For quantized particles, the physical meaning of the parameter $p$ would
be $p_3/\hbar$. However, the present argument is strictly classical
(quantization will be discussed later). The Fourier transform of
Eq.~(\ref{KG}) is
\begin{equation}
-2ip\,\gamma^{03}\,\chi_{,0}-ip\,\gamma^{03}{}_{,0}\,\chi
-p^2\,\gamma^{33}\,\chi-
2ip\,\gamma^{m3}\,\chi_{,m}-ip\,\gamma^{m3}{}_{,m}\,\chi+
(\gamma^{mn}\,\chi_{,m})_{,n}=-\kappa^2\,\gamma\,\chi.
\end{equation}
Let us further define \,$\psi\equiv\sqrt{\gamma^{03}}\,\chi$. After some
rearrangement, Eq.~(\theequation) can be written as
\begin{equation}
i\,\partial\psi/\partial u=
\mit\Omega(u,x,y,p,\partial_x,\partial_y)\,\psi, \label{Ham}
\end{equation}
where $\mit\Omega$ is a linear operator acting on the space of complex
functions $\psi(u,x,y,p)$:
\begin{equation}\begin{array}{l}
{\mit\Omega}\,\psi=
\{(\kappa^2\gamma-p^2\gamma^{33})\,\psi/\sqrt{\gamma^{03}}+
[\gamma^{mn}\,(\psi/\sqrt{\gamma^{03}}\,)_{,m}]_{,n}\\
  \hspace*{20mm} -\,2ip\,\gamma^{m3}\,(\psi/\sqrt{\gamma^{03}}\,)_{,m}
-ip\,\gamma^{m3}{}_{,m}\,(\psi/\sqrt{\gamma^{03}}\,)\}
/2p\sqrt{\gamma^{03}}.
\end{array}\end{equation}
We are using here a mathematical technique borrowed from ordinary
quantum mechanics, but at this stage the problem is completely
classical.

It will now be shown that $\int\psi^*{\mit\Omega}\psi\,dx\,dy$ is real
(that is, $\mit\Omega$ is a Hermitian operator) if $\psi$ falls rapidly
enough for large $x$ and~$y$. This result is obvious for the first two
terms in the braces on the right hand side of (\theequation). For the
third term, we have, after integrating by parts,
\begin{equation}
\int\psi^*\,[(\gamma^{mn}\,\chi_{,m})_{,n}/\sqrt{\gamma^{03}}\,]\,dx\,dy
=-\int\chi_{,n}^*\,\gamma^{mn}\,\chi_{,m}\,dx\,dy,
\end{equation}
which is also real.  Likewise, the last two terms together give
\begin{equation}
-(i/2)\int\chi^*\,(2\,\gamma^{m3}\,\chi_{,m}+\gamma^{m3}{}_{,m}\,\chi)
\,dx\,dy=-(i/2)\int(\chi^*\,\gamma^{m3}\,\chi_{,m}
-\chi^*_{,m}\,\gamma^{m3}\,\chi)\,dx\,dy,
\end{equation}
where an integration by parts was performed on the second term on the
left hand side. Again, the result is seen to be real.

It therefore follows from Eq.~(\ref{Ham}) that
\begin{equation}
(d/du)\,\int|\psi(u,x,y,p)|^2\,dx\,dy=0\qquad\forall\,p,
\end{equation}
so that the expression $\int|\psi(u,x,y,p)|^2\,dx\,dy$ is a constant of
the motion (it has the same value on all the hypersurfaces $u={\rm
const.})$. We thus have an infinity of conservation laws, one for each
value of $p$.

We now turn our attention to the quantized version of the above theory.
We assume that a quantum field $\mit\Psi$ satisfies the same
Klein-Gordon equation as the classical field, for the given classical
background $g_{\alpha\beta}$:
\begin{equation}
i\,\partial{\mit\Psi}(u,x,y,p)/\partial u=
\mit\Omega(u,x,y,p,\partial_x,\partial_y)\,{\mit\Psi}(u,x,y,p).
\label{HAM}
\end{equation}
Note that $\mit\Psi$ is a quantum field operator, acting on the Hilbert
space of quantum states, while $\mit\Omega$ is a purely classical linear
operator referring to the classical parameters $u, \,x, \,y$, and $p$.
The problem is whether Eq.~(\theequation) generates a {\em unitary\/}
transformation of the field operator $\mit\Psi$.

This question is not trivial, because the present classical field theory
does not have the usual canonical structure, and there is no guarantee
that quantization is possible [9]. The term ${1\over
2}\,\gamma^{\alpha\beta} \phi_{,\alpha}\phi_{,\beta}$ in the Lagrangian
density $\cal L$ produces a canonical momentum $\pi\equiv\partial{\cal
L}/\partial\phi_{,0} =\gamma^{03}\phi_{,3}\,$. This relationship is a
{\em constraint\/} restricting the values of $\pi$ and $\phi$ on a
constant $x^0$ hypersurface. Moreover, the Hamiltonian density, ${\cal
H}\equiv\pi\phi_{,0}-{\cal L}$, does not depend on $\pi$ at all! The
analogy between Poisson brackets and commutators therefore fails, and it
is not obvious that there is a unitary evolution law
$\,i\partial{\mit\Psi}(u,x,y,p)/\partial u= [{\mit\Psi}(u,x,y,p),H]$,
where $H$ is a Hermitian operator acting on the quantum Hilbert space.
This ought to be {\em proved.}

Since the classical $\psi$ satisfied \,$\psi(u,x,y,-p)=\psi^*(u,x,y,p)$,
we shall assume the same for the quantized field:
\begin{equation}
{\mit\Psi}^\dagger(u,x,y,p)\equiv{\mit\Psi}(u,x,y,-p), \label{psidag}
\end{equation}
where a dagger denotes the adjoint of an operator in the quantum Hilbert
space (an asterisk denotes the complex conjugate of an ordinary number).
We shall henceforth consider both ${\mit\Psi}$ and ${\mit\Psi}^\dagger$
as independent variables, and on the other hand we restrict $p$ to
positive values [2,3]. We further postulate the equal-$u$ commutation
relations (whose consistency we shall have to check, since they cannot
be derived from an analogy with classical Poisson brackets):
\begin{equation}
[{\mit\Psi}(u,x',y',p'),{\mit\Psi}(u,x'',y'',p'')]=0,\label{CR1}
\end{equation}
\begin{equation}
[{\mit\Psi}^\dagger(u,x',y',p'),{\mit\Psi}^\dagger(u,x'',y'',p'')]=0,
\end{equation}
\begin{equation}
[{\mit\Psi}(u,x',y',p'),{\mit\Psi}^\dagger(u,x'',y'',p'')]=
  \delta(x'-x'')\,\delta(y'-y'')\,\delta(p'-p''),\label{CR2}
\end{equation}
where an arbitrary normalization constant on the right hand side of
(\theequation) has been ignored.

Finally, we define, in the quantum Hilbert space, a Hermitian operator
$H$, which will play the role of a Hamiltonian, as we shall see:
\begin{equation}
H\equiv\int{\mit\Psi}^\dagger(u,x,y,p)\,
{\mit\Omega\,\Psi}(u,x,y,p)\,dx\,dy\,dp.
\end{equation}
We then indeed have
\begin{equation}
[{\mit\Psi}(u,x',y',p'),H]={\mit\Omega\,\Psi}(u,x',y',p'),
\end{equation}
and we can consistently postulate that both sides of (\theequation) are
equal to $i\,\partial{\mit\Psi}(u,x',y',p')/\partial u$, in agreement
with Eq.~(\ref{HAM}). This justifies the name ``Hamiltonian'' given to
the operator $H$ defined above. That operator generates a {\em
unitary\/} dynamical evolution: the commutation relations
(\ref{CR1}--\ref{CR2}) remain valid for all $u$.

This can also be seen in a more complicated but more general way, which
will be useful later, because it does not rely on the existence of a
Hamiltonian. We can directly differentiate Eq.~(\ref{CR2}) with respect
to $u$, and use Eq.~(\ref{HAM}) and its adjoint,
\begin{equation}
i\,\partial{\mit\Psi}^\dagger(u,x,y,p)/\partial u=
\tilde{\mit\Omega}(u,x,y,p,\partial_x,\partial_y)\,
{\mit\Psi}^\dagger(u,x,y,p),
\end{equation}
where $\tilde{\mit\Omega}(u,x,y,p,\partial_x,\partial_y)\equiv
{\mit\Omega}(u,x,y,-p,\partial_x,\partial_y)$, by virtue of
(\ref{psidag}). (We can also write this as \,$\tilde{\mit\Omega}=
-{\mit\Omega}^*$.) Since $\mit\Omega$ is an operator acting on functions
of $x, \,y$, and $p$, and is local in these coordinates, it is
straightforward (but tedious) to verify that
\begin{equation}
i\,(\partial/\partial u)\,[({\mit\Psi})',({\mit\Psi}^\dagger)'']=
[({\mit\Omega\,\Psi})',({\mit\Psi}^\dagger)'']+
[({\Psi})',(\tilde{\mit\Omega}\,{\mit\Psi}^\dagger)'']=0, \label{CR0}
\end{equation}
where $({\mit\Psi})'$ means ${\mit\Psi}(u,x',y',p')$, and likewise for
the other symbols. It is crucial in the verification of (\theequation)
that $\mit\Omega$ be both local and Hermitian.

The commutation relations (\ref{CR1}--\ref{CR2}) show that the operators
${\mit\Psi}$ and ${\mit\Psi}^\dagger$ behave as annihilation and
creation operators, respectively. In a flat spacetime, it is possible to
Fourier transform the coordinates $x$ and $y$ too, and it then becomes
manifest that the transformed operators refer to particles of mass
$\hbar\kappa/c$. [This will be explicitly shown later, see
Eq.~(\ref{aa}) and the ensuing discussion.]

Consider in particular the case of a ``sandwich wave'': for $u<u_1$ and
$u>u_2\,$, spacetime is flat ($g^{03}=-g^{11}=-g^{22}=1$, other
$g^{\alpha\beta}$ vanish). As we have just seen, there are {\em
separate\/} linear field equations for ${\mit\Psi}$ and for
${\mit\Psi}^\dagger$, even in the region $u_1<u<u_2$ (this is related to
the conservation of the classical momentum $p_3$). Nowhere is there any
mixing of creation and annihilation operators. Therefore there is no
Bogoliubov transformation. The number of particles is conserved, and in
particular no particles emerge from the vacuum [4--6]. The meaning of
the more general conservation laws that we found here is that {\em for
each value of $p_3$ separately\/}, the number of particles with momentum
$p_3$ is conserved.

Finally, we turn our attention to more general null-fronted metrics,
where $g^{0m}\neq 0$. In that case, the classical field equations can be
written as
\begin{equation}
A\,(i\,\partial\psi/\partial u)=B\,\psi, \label{AB}
\end{equation}
where $\psi\equiv\sqrt{\gamma^{03}}\,\chi$ as before, and where $A$ and
$B$ are linear operators acting on the complex function $\psi(u,x,y,p)$.
Explicitly,
\begin{equation}
A=2p+i\,[(\partial/\partial x^m)\,(\gamma^{0m}/\gamma^{03})+
(\gamma^{0m}/\gamma^{03})\,(\partial/\partial x^m)],
\end{equation}
which is manifestly Hermitian. Then, with the exception of a subspace of
functions satisfying $A\psi=0$, there exists [10] a unique inverse
operator, $A^{-1}$, and we can write
\begin{equation}
i\,\partial\psi/\partial u=(A^{-1}B)\,\psi.
\end{equation}
Note that $A^{-1}$ is not local (the inverse of a differential operator
is an integral operator) but this is not the main issue. The difficulty
is that $A^{-1}B$ is not, in general Hermitian. Consequently, if we
attempt to have a quantized field operator $\mit\Psi$ which obeys an
equation analogous to (\theequation), the commutation relation
(\ref{CR2}) is in jeopardy.

Let us substitute $A^{-1}B$ for $\mit\Omega$ in Eq.~(\ref{CR0}) (and
likewise $\tilde{A}^{-1}\tilde{B}$ for $\tilde{\mit\Omega}$, where a
tilde on $A$ or $B$ means the substitution $p\to-p$), and let us
multiply the result by $(A)'(\tilde{A})''$, to eliminate the
nonlocality. We obtain an expression
\begin{equation}
[(B{\mit\Psi})',(\tilde{A}{\mit\Psi}^\dagger)'']+
[(A{\mit\Psi})',(\tilde{B}{\mit\Psi}^\dagger)''], \label{LHS}
\end{equation}
and the question is whether this sum vanishes, so that (\ref{CR2}) can
hold for all $u$.

As a simple example, consider the case where the only nonvanishing
components of $\gamma^{\alpha\beta}$ are \,$\gamma^{03}=-\gamma^{11}=
-\gamma^{22}=1$, and \,$\gamma^{01}=f(u)$. Let $g(u)\equiv df/du$. We
have
\begin{equation}
A=2p+2i\,f(u)\,\partial/\partial x,
\end{equation}
\begin{equation}
B=-(\partial^2/\partial x^2)-(\partial^2/\partial y^2)+\kappa^2+
g(u)\,\partial/\partial x.
\end{equation}
Since the $\gamma^{\alpha\beta}$ do not depend on $x$ or $y$, we can
perform a Fourier transformation on these two coordinates, and define
\begin{equation}
a(u,{\bf k},p)\equiv(2\pi)^{-1}\int e^{-i{\bf k\cdot r}}\,
{\mit\Psi}(u,x,y,p)\,dx\,dy,
\end{equation}
where ${\bf k}\equiv\{k_x\,,k_y\}$ and ${\bf r}=\{x, y\}$. We likewise
define the adjoint operator $a^\dagger(u,{\bf k},p)$. It is
straightforward to show from Eq.~(\ref{CR2}) that
\begin{equation}
[a(u,{\bf k}',p'),a^\dagger(u,{\bf k}'',p'')]
=\delta({\bf k}'-{\bf k}'')\,\delta(p'-p'').
\label{aa}\end{equation}
The Fourier transform of the field equation (\ref{AB}) is, with the
values of $A$ and $B$ specified above,
\begin{equation}
2\,(p-f\,k_x)\,(i\,\partial a/\partial u)=
({\bf k}^2+\kappa^2+i\,g\,k_x)\,a,
\end{equation}
and likewise for $a^\dagger$.

In the special case of flat space ($f=g=0$), Eq.~(\theequation) reduces
to the equation of motion (in the Heisenberg picture) for the
annihilation operator, $a(u,{\bf k},p)$, of a particle of mass
$\hbar\kappa/c$. However, if $g\neq 0$, the situation is more intricate.
Let us assume that the commutation relation (\ref{aa}) is valid at some
``time'' $u_1\,$. At later times, we have, from (\theequation),
\begin{equation}
a(u,{\bf k},p)=a(u_1,{\bf k},p)\;Z(u_1\,,u,{\bf k},p), \label{aaZ}
\end{equation}
where
\begin{equation}
Z(u_1\,,u,{\bf k},p)=\exp\left(-i\int^u_{u_1}\frac{{\bf k}^2
+\kappa^2+ig(u')\,k_x}{2\,[\,p-f(u')\,k_x\,]}\,du'\right).
\end{equation}
Because of the $ig$ term in the numerator, $|Z|\neq 1$ and
Eq.~(\ref{aaZ}) is not a unitary transformation. A straightforward
calculation gives
\begin{equation}
[a(u,{\bf k}',p'),a^\dagger(u,{\bf k}'',p'')]
=\delta({\bf k}'-{\bf k}'')\,\delta(p'-p'')\;|Z(u_1\,,u,{\bf k}',p')|^2.
\end{equation}
If this is transformed back into coordinate space, the result is not
proportional to \mbox{$\delta(x'-x'')\,\delta(y'-y'')$,} but is
\begin{equation}
[({\mit\Psi})',({\mit\Psi}^\dagger)'']=(2\pi)^{-1}\,\delta(p'-p'')
\,\int|Z(u_1\,,u,{\bf k},p)|^2\,
e^{i{\bf k}\cdot({\bf r}'-{\bf r}'')}\,d{\bf k}.
\end{equation}
This is a {\em nonlocal\/} function of $(x'-x'')$ and $(y'-y'')$.
The commutation relation (\ref{CR2}) is therefore incompatible with the
field equation (\ref{HAM}). The quantum field ${\mit\Psi}(u,x,y,p)$
cannot be interpreted in terms of particle creation and annihilation
operators, and the physical meaning of the theory is problematic. [In
this particular case, there is a simple way out of the dilemma:
returning to the usual time, $t=(u+v)/\sqrt{2}$. Surfaces of constant
$t$ have everywhere a time-like normal vector, and a Cauchy problem can
be properly formulated. However, this may not be possible for more
general metrics.]

\bigskip Work by OL was supported by the Technion Graduate School. Work
by AP was supported by the Gerard Swope Fund and the Fund for
Encouragement of Research.\bigskip 

\frenchspacing
\begin{enumerate}
\item P. A. M. Dirac, {\it Rev. Mod. Phys.\/} {\bf 21}, 392 (1949).
\item H. Leutwyler, J. R. Klauder, and L. Streit, {\em Nuovo Cimento A}
{\bf 66}, 536 (1970).
\item F. Coester, {\em Prog. Part. Nucl. Phys.\/} {\bf 29}, 1 (1992).
\item S. Deser, {\em J. Phys. A} {\bf 8}, 1972 (1975).
\item G. W. Gibbons, {\em Comm. Math. Phys.\/} {\bf 45}, 191 (1975).
\item J. Garriga and E. Verdaguer, {\em Phys. Rev. D} {\bf 43}, 391
(1991).
\item A. Peres, {\em Phys. Rev. Letters\/} {\bf 3}, 571 (1959).
\item A. Peres, {\em Phys. Rev.\/} {\bf 118}, 1105 (1960).
\item S. A. Hojman and L. C. Shepley, {\em J. Math. Phys.\/} {\bf 32},
142 (1991).
\item F. Riesz F and B. Sz.-Nagy, {\it Functional Analysis\/}, Ungar,
New York (1955) p.~152.
\end{enumerate}\newpage\begin{verbatim}

                 QUANTUM THEORY: CONCEPTS AND METHODS

                           by Asher Peres

          446 + xiv pages                 ISBN 0-7923-2549-4

in USA and Canada:                  in all other countries:

Kluwer Academic Publishers          Kluwer Academic Publishers
P. O. Box 358                       P. O. Box 322
Accord Station                      3300 AH Dordrecht
Hingham, MA 02018-0358              The Netherlands

Phone 617-871-6600                  Phone (0)78 - 524400
Fax   617-871-6528                  Fax   (0)78 - 524474


                         TABLE OF CONTENTS

Chapter 1: Introduction to Quantum Physics

1-1. The downfall of classical concepts                             3
1-2. The rise of randomness                                         5
1-3. Polarized photons                                              7
1-4. Introducing the quantum language                               9
1-5. What is a measurement?                                        14
1-6. Historical remarks                                            18
1-7. Bibliography                                                  21

Chapter 2: Quantum Tests

2-1. What is a quantum system?                                      24
2-2. Repeatable tests                                               27
2-3. Maximal quantum tests                                          29
2-4. Consecutive tests                                              33
2-5. The principle of interference                                  36
2-6. Transition amplitudes                                          39
2-7. Appendix: Bayes's rule of statistical inference                45
2-8. Bibliography                                                   47

\end{verbatim}\newpage\begin{verbatim}
Chapter 3: Complex Vector Space

3-1. The superposition principle                                    48
3-2. Metric properties                                              51
3-3. Quantum expectation rule                                       54
3-4. Physical implementation                                        57
3-5. Determination of a quantum state                               58
3-6. Measurements and observables                                   62
3-7. Further algebraic properties                                   67
3-8. Quantum mixtures                                               72
3-9. Appendix: Dirac's notation                                     77
3-10. Bibliography                                                  78

Chapter 4: Continuous Variables

4-1. Hilbert space                                                  79
4-2. Linear operators                                               84
4-3. Commutators and uncertainty relations                          89
4-4. Truncated Hilbert space                                        95
4-5. Spectral theory                                                99
4-6. Classification of spectra                                     103
4-7. Appendix: Generalized functions                               106
4-8. Bibliography                                                  112

Chapter 5: Composite Systems

5-1. Quantum correlations                                          115
5-2. Incomplete tests and partial traces                           121
5-3. The Schmidt decomposition                                     123
5-4. Indistinguishable particles                                   126
5-5. Parastatistics                                                131
5-6. Fock space                                                    137
5-7. Second quantization                                           142
5-8. Bibliography                                                  147

Chapter 6: Bell's Theorem

6-1. The dilemma of Einstein, Podolsky, and Rosen                  148
6-2. Cryptodeterminism                                             155
6-3. Bell's inequalities                                           160
6-4. Some fundamental issues                                       167
6-5. Other quantum inequalities                                    173
6-6. Higher spins                                                  179
6-7. Bibliography                                                  185

Chapter 7: Contextuality

7-1. Nonlocality versus contextuality                              187
7-2. Gleason's theorem                                             190
7-3. The Kochen-Specker theorem                                    196
7-4. Experimental and logical aspects of inseparability            202
7-5. Appendix: Computer test for Kochen-Specker contradiction      209
7-6. Bibliography                                                  211

Chapter 8: Spacetime Symmetries

8-1. What is a symmetry?                                           215
8-2. Wigner's theorem                                              217
8-3. Continuous transformations                                    220
8-4. The momentum operator                                         225
8-5. The Euclidean group                                           229
8-6. Quantum dynamics                                              237
8-7. Heisenberg and Dirac pictures                                 242
8-8. Galilean invariance                                           245
8-9. Relativistic invariance                                       249
8-10. Forms of relativistic dynamics                               254
8-11. Space reflection and time reversal                           257
8-12. Bibliography                                                 259

Chapter 9: Information and Thermodynamics

9-1. Entropy                                                       260
9-2. Thermodynamic equilibrium                                     266
9-3. Ideal quantum gas                                             270
9-4. Some impossible processes                                     275
9-5. Generalized quantum tests                                     279
9-6. Neumark's theorem                                             285
9-7. The limits of objectivity                                     289
9-8. Quantum cryptography and teleportation                        293
9-9. Bibliography                                                  296

Chapter 10: Semiclassical Methods

10-1. The correspondence principle                                 298
10-2. Motion and distortion of wave packets                        302
10-3. Classical action                                             307
10-4. Quantum mechanics in phase space                             312
10-5. Koopman's theorem                                            317
10-6. Compact spaces                                               319
10-7. Coherent states                                              323
10-8. Bibliography                                                 330

Chapter 11: Chaos and Irreversibility

11-1. Discrete maps                                                332
11-2. Irreversibility in classical physics                         341
11-3. Quantum aspects of classical chaos                           347
11-4. Quantum maps                                                 351
11-5. Chaotic quantum motion                                       353
11-6. Evolution of pure states into mixtures                       369
11-7. Appendix: PostScript code for a map                          370
11-8. Bibliography                                                 371

Chapter 12: The Measuring Process

12-1. The ambivalent observer                                      373
12-2. Classical measurement theory                                 378
12-3. Estimation of a static parameter                             385
12-4. Time-dependent signals                                       387
12-5. Quantum Zeno effect                                          392
12-6. Measurements of finite duration                              400
12-7. The measurement of time                                      405
12-8. Time and energy complementarity                              413
12-9. Incompatible observables                                     417
12-10. Approximate reality                                         423
12-11. Bibliography                                                428

Author Index                                                       430

Subject Index                                                      435
\end{verbatim}
\end{document}